\documentclass[english]{revtex4-2}
\usepackage[T1]{fontenc}
\usepackage[latin9]{inputenc}
\usepackage{babel}
\usepackage{amsmath}
\usepackage{amssymb}
\usepackage{graphicx}
\usepackage[unicode=true,pdfusetitle,
 bookmarks=true,bookmarksnumbered=false,bookmarksopen=false,
 breaklinks=false,pdfborder={0 0 1},backref=false,colorlinks=false]
 {hyperref}
\begin{document}
\title{Universality and hysteresis in slow sweeping of bifurcations}
\author{Roie Ezraty(1), Ido Levin(1,2), and Omri Gat(1)}
\address{(1) Racah Institute of Physics, Hebrew University of Jerusalem, Jerusalem,
Israel 9190401~\\
(2) Department of Chemistry, University of Washington, Seattle, WA,
USA}
\begin{abstract}
Bifurcations in dynamical systems are often studied experimentally
and numerically using a slow parameter sweep. Focusing on the cases
of period-doubling and pitchfork bifurcations in maps, we show that
the adiabatic approximation always breaks down sufficiently close
to the bifurcation, so that the upsweep and downsweep dynamics diverge
from one another, disobeying standard bifurcation theory. Nevertheless,
we demonstrate universal upsweep and downsweep trajectories for sufficiently
slow sweep rates, revealing that the slow trajectories depend essentially
on a structural asymmetry parameter, whose effect is negligible for
the stationary dynamics. We obtain explicit asymptotic expressions
for the universal trajectories, and use them to calculate the area
of the hysteresis loop enclosed between the upsweep and downsweep
trajectories as a function of the asymmetry parameter and the sweep
rate.
\end{abstract}
\maketitle

\section{Introduction}

A key property of dissipative dynamical systems is that bounded trajectories
converge towards an attractor, whose properties thus determine the
persistent long-term behavior of the system. When the dynamical system
depends smoothly on a parameter, the attractor varies smoothly as
a function of the parameter, except at special parameter values, where
a small parameter variation induces a sharp transition---a bifurcation---in
the long-time dynamics of the system. The study of bifurcations is
facilitated by the fact that their local structure is universal and
captured by a normal form dynamical system whose dimension is equal
to the codimension of the bifurcation \citep{wiggins2013global}.

Here we study the persistent dynamics of systems whose parameters
are time-dependent and change \emph{adiabatically, }in the sense that
the time scale of parameter variation is much longer than any dynamical
time scale. When the system parameters are far from bifurcation points,
the convergence to the attractor occurs on a dynamical time scale,
and therefore the trajectory of the non-autonomous system follows
the time-dependent attractor of the autonomous system with the instantaneous
(also called frozen) parameter values, up to an error that tends to
zero as the adiabatic time scale tends to infinity. However, bifurcations
typically involve the exchange of stability, so that the time scale
of convergence to the attractor diverges when the system parameters
approach a bifurcation point. Therefore, no matter how slowly the
parameter is varied, the variation is no longer adiabatic close enough
to the bifurcation. Does this observation imply that parameter sweeps
are useless as probes of bifurcations? Not necessarily, but the dynamics
near a swept bifurcation are qualitatively different than that of
an autonomous dynamical system. In particular, the dynamics depend
on the direction of the sweep and exhibit hysteresis.

In the context of continuous-time dynamics, this phenomenon, which
has been dubbed ``la chasse au canard'' (i.e., duck hunting) \citep{benoit1981chasse},
has been studied quite thoroughly, both numerically and asymptotically
using singular perturbation theory, see e.g.~\citep{mandel1987slow,szmolyan2001canards}.
Here we focus on the less-studied case of adiabatic parameter sweep
of bifurcations of discrete-time dynamical systems, specifically of
period-doubling and the closely related pitchfork bifurcations. In
experiments, the issue of bifurcations in discrete-time arises naturally
when studying periodically driven systems, like AC-driven nonlinear
electric circuits, which are sampled at drive-period intervals \citep{klinker1984period,morris1986postponed,pieranski1987noise,liauw1993periodic,Linsay1981}.
In these cases, parameter sweeping is often used to efficiently extract
the attractor for a range of parameter values. The breakdown of adiabaticity
and hysteresis in maps undergoing period-doubling was demonstrated
numerically in \citep{morris1986postponed,pieranski1987noise,kapral1985bifurcation}.
A theory of the universal dynamics of maps in the vicinity of a period-doubling
bifurcation, based on singular perturbation theory and asymptotic
analysis was first put forward in \citep{baesens1991slow}, and further
developed in \citep{baesens1995gevrey,davies1997dynamic}. These works
introduced the concept of the adiabatic manifold, an attracting submanifold
of the extended phase space, on which the dynamics are slow. The theory
of \citep{baesens1991slow,baesens1995gevrey,davies1997dynamic} is
strong and general, but difficult to apply to specific examples, and
the adiabatic manifold approach has not yet been used in an experimental
context.

This work aims to derive theoretical results that are directly applicable
to experiments and simulations of slow sweeping of period-doubling
bifurcations. For this purpose, we analyze asymptotic trajectories
that start sufficiently far from the bifurcation. These trajectories
converge on the instantaneous autonomous system attractor well before
the control parameter approaches the bifurcation region, facilitating
the application of the universal normal form to a physical system
of interest. Moreover, we use the resulting effective dynamics to
derive several new results, most notably the shape and area of the
hysteresis loop in the bifurcation diagram, and its dependence on
the adiabatic small parameter, $\varepsilon$, (the sweep rate of
the control parameter) that can be easily measured in an experiment
or numerical simulation of a given system.

Specifically, we study the trajectory of a one-degree-of-freedom map
depending on a single parameter that is adiabatically swept through
a supercritical period-doubling bifurcation of a fixed point, by considering
the trajectory of its second iterate, which has fixed point attractors
on both sides of the bifurcation, in section \ref{subsec:The-normal-form-map}.
The slow sweep implies that the fixed points vary slightly between
subsequent iterations of the map, allowing us to approximate the map
by a one-degree-of-freedom continuous-time flow that undergoes a supercritical
pitchfork bifurcation, as shown in section \ref{subsec:Continuous-time-dynamics-approxi}.
We next study the dynamics near the bifurcation using the pitchfork
normal form, showing that the universal adiabatic trajectories depend
strongly on the \emph{asymmetry parameter} $w$ that inflicts an overall
drift of the fixed points near the bifurcation; the effect of this
parameter is negligible in autonomous dynamics. In sections \ref{subsec:Adiabatic-upsweep}
and \ref{subsec:Adiabatic-downsweep} we solve the normal-form equations
of motion, and calculate the upsweep and downsweep trajectories explicitly
in the limit where $\varepsilon$, the adiabatic parameter is small.
We observe that a key role is played by the parameter $s=\varepsilon^{1/4}w$,
which determines the \emph{shape} of the trajectories, in the sense
that adiabatic trajectories with the same $s$ value are related by
a scaling transformation. For a fixed map, the shape parameter tends
to zero in the adiabatic limit, but the upsweep trajectory does not
approach a limiting form. As a consequence, the area of the hysteresis
loop has a complicated behavior in the adiabatic limit: its leading
asymptotic is $\frac{2}{3}\varepsilon^{1/4}\left(-\log\left(2\pi s^{2}\sqrt{-\log\left(2\pi s^{2}\right)}\right)\right)^{3/4}$,
as shown in section \ref{subsec:The-hysteresis-loop}. The results
outlined so far were derived for the adiabatic normal form map; in
section \ref{subsec:Logistic-map} we use the normal form theory to
calculate the adiabatic sweep trajectories of the logistic map in
the region of the fundamental period-doubling bifurcation, demonstrating
the universality of our theory. Finally, we present our conclusions
in section \ref{sec:Discussion-and-Conclusions}.

\section{The effective dynamics}

\subsection{The normal-form adiabatic sweep map\label{subsec:The-normal-form-map}}

\begin{figure}[t]
\includegraphics[scale=0.7]{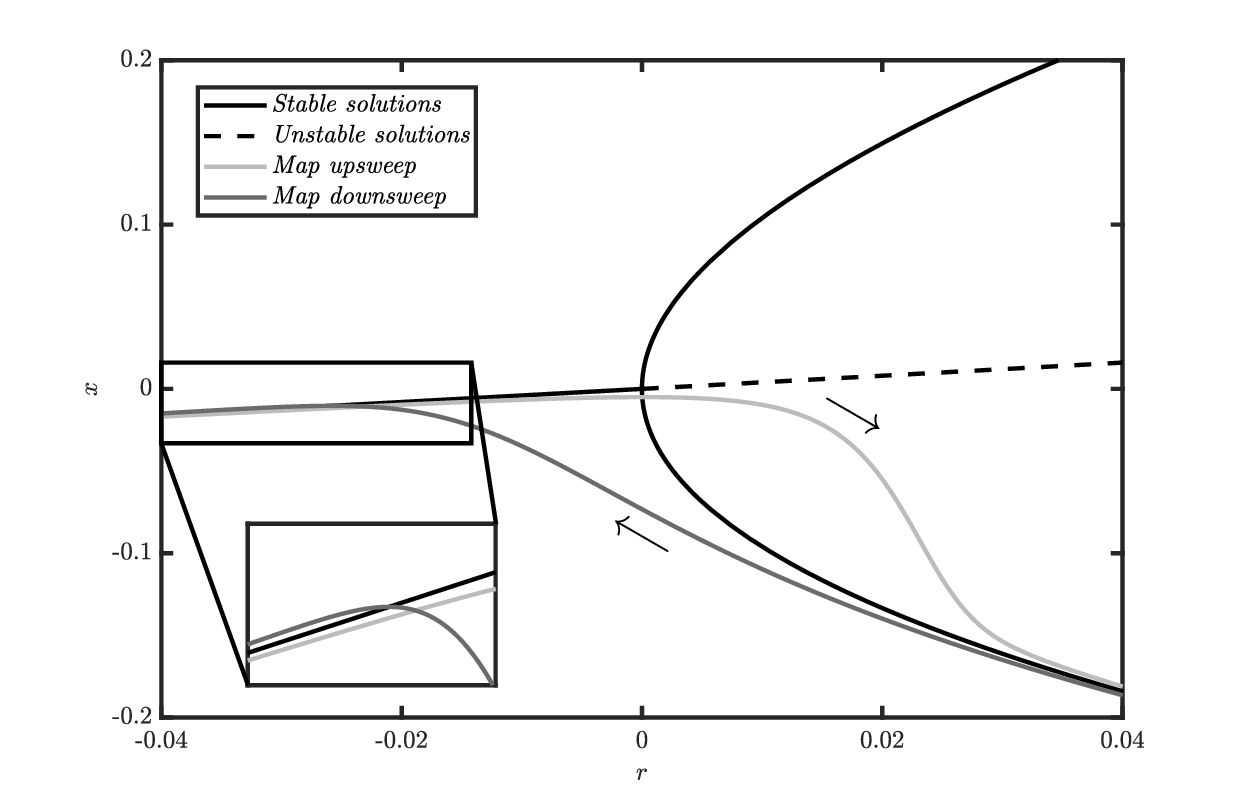}

\caption{\label{fig:Map and stable sols}Dynamics of the map given in equation
(\ref{eq:pitchfork}) with $\varepsilon=10^{-4}$ and $s=0.04$ ($w=0.4$)
in pale (upweep) and darker (downsweep) gray. Black lines represent
the stable fixed points at $r<0$ (single point) and $r>0$ (two solutions
for every $r$), where at $r=0$ a pitchfork bifurcation occurs. The
inset is a magnification of the parts of the trajectories that are
close to $y_{0}$, demonstrating that the trajectories can overshoot
the stable fixed point and cross each other. Moreover, even though
the dynamics are initiated exactly at the stable fixed point at $t<<0$,
they chase the fixed point from below due to the positive sign of
$w$, hence after the bifurcation at $t=0$ the cynamics always converges
to the lower branch stable fixed point.}
\end{figure}

We consider non-autonoumous dynamics generated by a time-dependent
one-degree-of-freedom map
\begin{equation}
x_{n+1}=F_{n}\left(x_{n}\right)\ ,\label{eq:adiamap}
\end{equation}
with $x,\,F_{n}$ real, $n$ integer. We assume that for a range of
$n$ values around zero, $F_{n}\left(x_{n}\right)$ has a fixed point
that undergoes a supercritical pitchfork bifurcation at $n=0$. This
means that for $n<0$, $F_{n}$ has a stable fixed point $y_{0}\left(n\right)$,
$F_{n}\left(y_{0}\left(n\right)\right)=y_{0}\left(n\right)$; at $n=0$,
$y_{0}$ becomes marginally stable with $\partial_{x}F_{0}\left(y_{0}\left(0\right)\right)=1$,
and for $n>0$, $y_{0}$ becomes unstable, and two new, stable fixed
points $y_{\pm}\left(n\right)$ appear (figure \ref{fig:Map and stable sols}).
The analysis of these dynamics covers also the case of adiabatic sweeping
through a period-doubling bifurcation, because when a map $G$ period-doubles,
its second iterate $F\equiv G^{2}$ undergoes a pitchfork bifurcation\citep{wiggins2013global}.
It is preferable to start with the adiabatic analysis of $F$ because,
unlike $G$, it has a stable fixed point for every $n$, facilitating
the continuum model approximation of section \ref{subsec:Continuous-time-dynamics-approxi}
below.

The normal form of a family of maps depending on the control parameter
$r$, which undergoes a pitchfork bifurcation at $r=0$, is \citep{wiggins2013global}
\begin{equation}
x_{n+1}=\left(1+r\right)x_{n}-x_{n}^{3}\ ,
\end{equation}
with fixed points $y_{0}=0$ and, for $r>0$, $y_{\pm}=\pm\sqrt{r}$.
When the dynamics is autonomous, $y_{0}$ can always be shifted to
zero by a change of coordinates, but for the adiabatic dynamics, $y_{0}$
varies with the parameter and its time dependence of must be taken
into account; for this reason we have to use a two-parameter normal
form map
\begin{equation}
x_{n+1}=u+\left(1+r\right)\left(x_{n}-u\right)-\left(x_{n}-u\right)^{3}
\end{equation}
with $y_{0}=u$. For adiabatic dynamics we let the parameters depend
on time so that

\begin{equation}
F_{n}\left(x_{n}\right)=u_{n}+\left(1+r_{n}\right)\left(x_{n}-u_{n}\right)-\left(x_{n}-u_{n}\right)^{3}\ .\label{eq:pitchfork}
\end{equation}

Adiabatic sweep means that the time scale of variation of the parameters
$r$ and $u$ is much longer than a single time step of the map; for
this reason, in the vicinity of the bifurcation it is sufficient to
approximate $r_{n}$, $u_{n}$ by linear functions. Since by shifting
$x$ and $n$ we can make $r_{0}=u_{0}=0$, we let

\begin{equation}
r_{n}=\pm\varepsilon n\,,\qquad u_{n}=\varepsilon wn\,,\label{eq:adiapar}
\end{equation}
where $0<\varepsilon\ll1,$ and $w$ is an order-one real number that
parametrizes the overall drift of the fixed points of the system.
Equations (\ref{eq:adiamap}), (\ref{eq:pitchfork}), (\ref{eq:adiapar}),
define the normal-form adiabatic sweep model. A positive (negative)
sign in (\ref{eq:adiapar}) models an adiabatic up(down)-sweep, respectively.

We can now define adiabatic upsweep and downsweep trajectories $x_{\uparrow n}$
and $x_{\downarrow n}$ (respectively) of the map $F$ that start
at some negative $n$ with sufficiently large absolute value, approach
one of the stable instantaneous fixed points of $F$, and are then
followed until $n$ becomes positive and large. Figure \ref{fig:Map and stable sols}
shows example upsweep and downsweep trajectories of the map (\ref{eq:pitchfork})
with $\varepsilon=10^{-4},\,w=0.4$, overlayed with the instantaneous
fixed points of the map. The sign of $w$ breaks the sign-flip symmetry
of the pitchfork bifurcation, and in the positive-$w$ case shown
in figure \ref{fig:Map and stable sols}, the upsweep trajectory remains
below $y_{0}$, eventually converging to $y_{-}$, which is used as
the starting point of the downsweep trajectory. Our next goal is to
study the adiabatic sweep trajectories and the hysteresis loop using
a continuum approximation.

\subsection{Continuous-time dynamics approximation\label{subsec:Continuous-time-dynamics-approxi} }

\begin{figure}[t]
\includegraphics[scale=0.65]{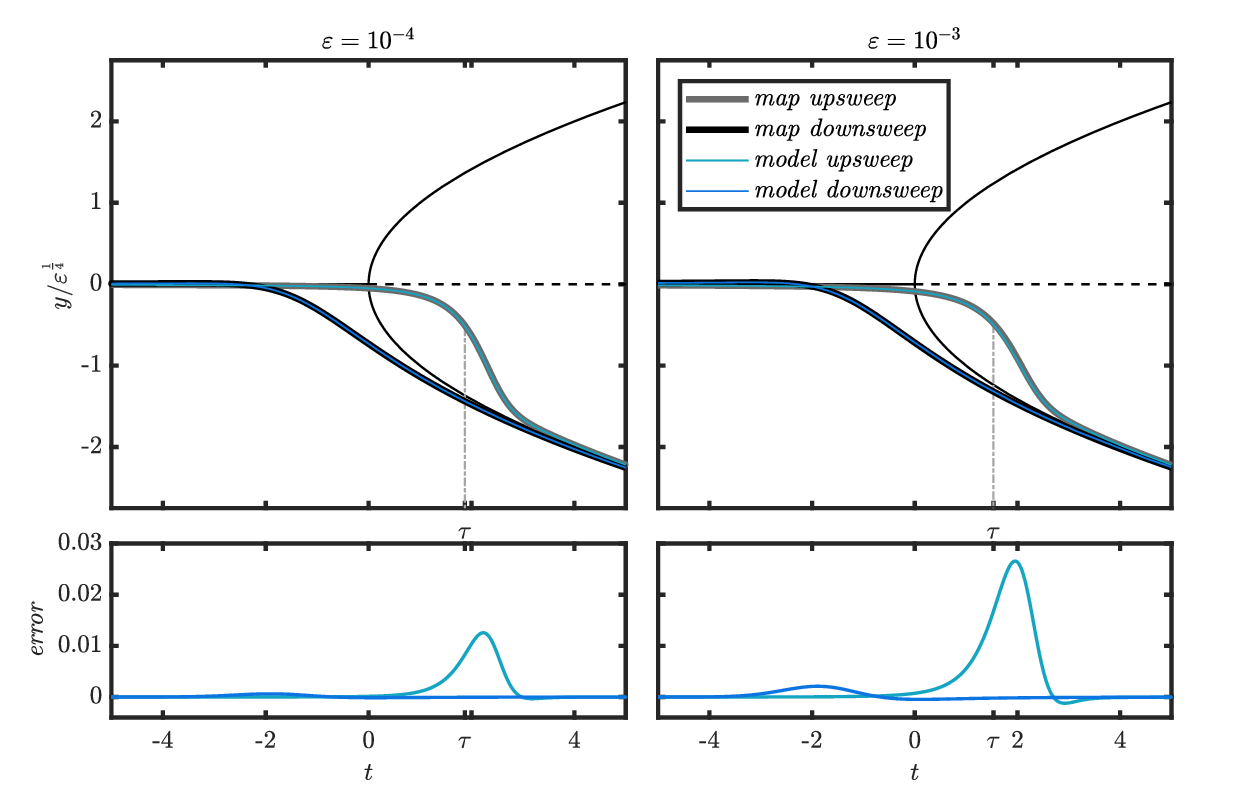}

\caption{\label{fig:Numerical model and map}Top: continuum model trajectories
from numerical solutions of equation (\ref{eq:nf-css}) in turquoise
(upsweep) and blue (downsweep) overlayed on the trajectories of the
map (\ref{eq:pitchfork}) (after transformation from $x$ and $n$
to $t$ and $y/\varepsilon^{1/4}$) in pale (upsweep) and dark (downsweep)
gray, alongside the fixed points of the map in filled (stable) and
dashed (unstable) black. Bottom: The difference between the trajectories
of the continuum model and the map, showing that as $\varepsilon\rightarrow0$
the trajectories of the continuous- and discrete-time systems approach
each other. Here we used $\varepsilon=10^{-4}\,,\,w=0.4$ ($s=0.04$)
in the left panels, and $\varepsilon=10^{-3}\,,\,w=0.4$ ($s=0.07$),
in the right panels.}
\end{figure}

The continuum model approximation starts from the observation that
since the parameters of the normal form map $F_{n}$ change slowly,
its fixed points change slowly, and therefore if $x_{n}$ is close
to a fixed point of $F_{n}$, then $x_{n+1}$ is close both to $x_{n}$
and to a fixed point of $F_{n+1}$. It follows that for trajectories
starting close to a fixed point, we can think of $n$ as a real variable
and approximate $x_{n+1}-x_{n}\sim dx/dn$, obtaining

\begin{equation}
\frac{dx}{dn}=\pm\varepsilon n\left(x_{n}-\varepsilon wn\right)-\left(x_{n}-\varepsilon wn\right)^{3}\ ,\label{eq:nf-c}
\end{equation}
that is simplified by letting $y_{n}=x_{n}-\varepsilon wn$ to

\begin{equation}
\frac{dy}{dn}=\pm\varepsilon ny_{n}-y_{n}^{3}-\varepsilon w\ ;\label{eq:nf-cs}
\end{equation}
as before a positive (negative) sign choice in the equation of motion
corresponds to upsweep (downsweep) dynamics, respectively. From the
linear term, the time scale in the adiabatic limit is $\sqrt{\varepsilon}$,
so we change the time variable to $t=\sqrt{\varepsilon}n$, obtaining
\begin{equation}
\frac{dy}{dt}=\pm ty-\frac{y^{3}}{\sqrt{\varepsilon}}-\sqrt{\varepsilon}w\ .\label{eq:nf-css}
\end{equation}
Figure \ref{fig:Numerical model and map} shows a comparison between
the numerical simulation of the continuous-time equation (\ref{eq:nf-css})
and the discrete time map dynamics in equation (\ref{eq:pitchfork}),
confirming the validity of the continuum model approximation.

Next, we want to identify the characteristic scale of $y$ for small
$\varepsilon$. It turns out that there are two regimes in (\ref{eq:nf-css})
with different characteristic scales. When $t<0$ for the upsweep
and $t>0$ for the downsweep trajectory, $y$ is attracted to the
instantaneous fixed point at 0, so that $y\sim\sqrt{\varepsilon}$
, and the cubic term is negligible. In this time interval, the appropriate
scaling is $y=\sqrt{\varepsilon}y_{s}$, where the subscript $_{s}$
indicates that $y$ approaches $y_{0}=0$, and the equation of motion
becomes
\begin{equation}
\frac{dy_{s}}{dt}=\pm ty_{s}-w-\sqrt{\varepsilon}y_{s}^{3}\ .\label{eq:nf-n}
\end{equation}
On the other hand, when $\pm t>0$, $y$ is attracted to one of the
nonzero instantaneous fixed points at $\pm\varepsilon^{1/4}\sqrt{\left|t\right|}$,
so that $y\sim\varepsilon^{1/4}$, and the term proportional to $w$
is negligible. In this time interval, the appropritate scaling is
$y=\varepsilon^{1/4}y_{l}$, where the subscript $_{l}$ indicates
that $y$ approaches a non-zero stable fixed point, and the equation
of motion becomes
\begin{equation}
\frac{dy_{l}}{dt}=\pm ty_{l}-y_{l}^{3}-\varepsilon^{1/4}w\ .\label{eq:nf-p}
\end{equation}
The crossover between the two regimes occurs at different times for
the upsweep and downsweep trajectories which we will now identify,
and then use to calculate the crossover trajectories.

The form (\ref{eq:nf-p}) of the effective dynamics equation of motion
makes it clear that the only dimensionless parameter in the adiabatic
dynamics is $s=\varepsilon^{1/4}w$; since $s$ determines the trajectories
$y\left(t\right)$ up to scaling, we will refer to it as the \emph{shape
parameter}. Note however that unlike $w$, $s$ tends to zero in the
adiabatic limit, so that the limit $\varepsilon\to0$ with $s$ fixed
is not appropriate for studying the adiabatic sweep problem. However,
this is a singular limit, so that the adiabatic trajectories and the
hysteresis loop do not converge to a well-defined shape.

\section{The adiabatic trajectories}

We seek solutions of the adiabatic equations of motion (\ref{eq:nf-css})
or their equivalents (\ref{eq:nf-n}), (\ref{eq:nf-p}), that start
far from the bifurcation at some $t<0$, and end far on the other
side of the bifurcation for $t>0$. Far from the bifurcation, the
adiabatic sweep follows a stable instantaneous fixed point up to small
deviations, so we seek solutions that start at these points for sufficiently
negative $t$, studying upsweep and downsweep trajectories separately.

\subsection{Adiabatic upsweep\label{subsec:Adiabatic-upsweep}}

Starting with $t<0$, $|t|$ of order one or larger, and $y_{s}$
of order one, the nonlinear term in equation (\ref{eq:nf-n}) is negligible.
In this approximation, the solution (choosing the positive sign for
upward sweep) that tends to 0 as $t\to-\infty$ is
\begin{equation}
y_{s\uparrow}\left(t\right)=-w\int_{-\infty}^{t}e^{\frac{1}{2}\left(t^{2}-\left(t'\right)^{2}\right)}dt'=-w\sqrt{\frac{\pi}{2}}e^{\frac{1}{2}t^{2}}\textrm{erfc\ensuremath{\left(-\frac{t}{\sqrt{2}}\right)}}\ ,\label{eq:ysup}
\end{equation}
where $\textrm{erfc}$ stands for the complementary error function
\citep{FIPS1402}. The approximation remains valid as long as $\left|y_{s\uparrow}\right|\ll\varepsilon^{-1/4}\Leftrightarrow\left|y_{\uparrow}\right|\ll\varepsilon^{1/4}$;
since $\textrm{erfc}\left(t\right)\sim e^{-t^{2}}/\left(\sqrt{\pi}t\right)$
as $t\to-\infty$, it follows that (\ref{eq:ysup}) remains valid
for all negative $t$, and since $\textrm{erfc}\left(t\right)\ensuremath{\to2}$
as $t\to-\infty$, also for positive $t$ such that $e^{\frac{1}{2}t^{2}}\ll\varepsilon^{-1/4}.$
For times later than this range, the nonlinear term in equation (\ref{eq:nf-css})
becomes important, so that the scaling of equation (\ref{eq:nf-p})
is applicable, while the term proportional to $w$ becomes negligible.
When this term is neglected, equation (\ref{eq:nf-p}) becomes a Bernoulli
equation, whose general solution is
\begin{equation}
y_{l\uparrow}\left(t\right)=\pm\frac{e^{t^{2}/2}}{\sqrt{c_{1}+\sqrt{\pi}\textrm{erfi}\left(t\right)}}\ ,\label{eq:ylup}
\end{equation}
where $\textrm{erfi}\left(t\right)=-i\left(1-\textrm{erfc}\left(it\right)\right)$,
and $c_{1}$ is an arbitrary constant of integration. For $t$ large
and positive, $\textrm{erfi\ensuremath{\left(t\right)}}\sim e^{t^{2}}/\left(\sqrt{\pi}t\right)$,
so that $y_{l}\left(t\right)\to\pm\sqrt{t}$ for all values of $c_{1}$;
the constant has to be determined by matching the right-hand side
of equation (\ref{eq:ylup}) with that of equation (\ref{eq:ysup}).

For this purpose we note that $w$ is negligible in equation (\ref{eq:nf-p})
as long as $|y_{l\uparrow}|\gg\varepsilon^{1/4}\Leftrightarrow|y_{\uparrow}|\gg\sqrt{\varepsilon}$,
so that there is an overlap interval $\sqrt{\varepsilon}\ll y_{\uparrow}\ll\varepsilon^{1/4}$,
where both equation (\ref{eq:ysup}) and equation (\ref{eq:ylup})
are valid. In this interval $y_{s\uparrow}\left(t\right)\sim-\sqrt{2\pi}we^{\frac{1}{2}t^{2}}$
and in order for this asymptote to match equation (\ref{eq:ylup}),
we must choose $c_{1}\gg\textrm{erfi}\left(t\right)$ in the interval
to ensure that $y_{l\uparrow}\left(t\right)\sim\pm c_{1}^{-1/2}e^{\frac{1}{2}t^{2}}$.
Recalling that $y_{s\uparrow}=\varepsilon^{-1/4}y_{l\uparrow}$, it
follows that the two asymptotes match if 
\begin{equation}
c_{1}=\left(2\pi w^{2}\right)^{-1}\varepsilon^{-1/2}
\end{equation}
and the ambiguous sign in (\ref{eq:ylup}) is chosen to match the
sign of $-w$.

In summary, we obtain the all-time aymptotic approximation for the
adiabatic upsweep trajectory
\begin{equation}
y_{\uparrow}\left(t\right)=\begin{cases}
-{\displaystyle \sqrt{\varepsilon}w\sqrt{\frac{\pi}{2}}e^{t^{2}/2}\textrm{erfc\ensuremath{\left(-\ensuremath{\frac{t}{\sqrt{2}}}\right)}}} & t\le0\,,\ \text{or}\ t>0\ \text{and}\;e^{\frac{1}{2}t^{2}}\ll\varepsilon^{-1/4}\\
-{\displaystyle \textrm{sign}\left(w\right)\frac{\varepsilon^{1/4}e^{t^{2}/2}}{\sqrt{\left(2\pi w^{2}\right)^{-1}\varepsilon^{-1/2}+\sqrt{\pi}\textrm{erfi}\left(t\right)}}} & t>0\ \text{and}\ 1\ll e^{\frac{1}{2}t^{2}}
\end{cases}\label{eq:upsweep}
\end{equation}
We check that the early- and late-time asymptotics agree in their
common interval of validity, and that $c_{1}\gg\textrm{erfi}\left(t\right)$
holds in this interval. It is useful to express the result also in
terms of the shape parameter $s=\varepsilon^{1/4}w$,

\begin{equation}
y_{\uparrow}\left(t\right)=-\varepsilon^{1/4}\begin{cases}
s\sqrt{{\displaystyle \frac{\pi}{2}}}e^{t^{2}/2}\textrm{erfc\ensuremath{\left(-\ensuremath{\frac{t}{\sqrt{2}}}\right)}} & t\le0\,,\ \text{or}\ t>0\ \text{and}\;e^{\frac{1}{2}t^{2}}\ll1/s\\
{\displaystyle \textrm{sign}\left(s\right)\frac{e^{t^{2}/2}}{\sqrt{\left(2\pi s^{2}\right)^{-1}+\sqrt{\pi}\textrm{erfi}\left(t\right)}}} & t>0\ \text{and}\ 1\ll e^{\frac{1}{2}t^{2}}
\end{cases}\label{eq:upsweep-s}
\end{equation}

\subsection{Adiabatic downsweep\label{subsec:Adiabatic-downsweep}}

Starting with $t<0$, $|t|$ of order one or larger, and $|y_{l}|$
comparable with $\sqrt{-t}$, the term proportional to $w$ in equation
(\ref{eq:nf-p}) is initially negligible. In this approximation, the
solutions of equation (\ref{eq:nf-p}) (choosing the negative sign
for downsweep) that approach $\pm\sqrt{-t}$ as $t\to-\infty$ are
\begin{equation}
y_{l\downarrow}\left(t\right)=\frac{\pm e^{-t^{2}/2}}{\pi^{1/4}\sqrt{\textrm{erfc\ensuremath{\left(-t\right)}}}}\ ,\label{eq:yldown}
\end{equation}
respectively. As in the case of equation (\ref{eq:ylup}), this approximation
remains valid as long as $|y_{l\downarrow}|\gg\varepsilon^{1/4}$,
which here holds for all negative $t$, and for positive $t$ small
enough that $e^{-t^{2}/2}\gg\varepsilon^{1/4}$. For later times we
can use equation (\ref{eq:nf-n}) without the nonlinear term, whose
general solution is 
\begin{equation}
y_{s\downarrow}\left(t\right)=e^{-t^{2}/2}\left(c_{2}-w\sqrt{\frac{\pi}{2}}\textrm{erfi}\left(\frac{t}{\sqrt{2}}\right)\right)
\end{equation}
valid when $|y_{s\downarrow}|\ll\varepsilon^{-1/4}$. In the overlap
interval $\sqrt{\varepsilon}\ll|y_{\downarrow}|\ll\varepsilon^{1/4}$,
$c_{2}\gg\textrm{erfi}\left(\frac{t}{\sqrt{2}}\right)$, and $\textrm{erfc\ensuremath{\left(t\right)}}\sim2$,
so that the asymptotes match if we choose 
\begin{equation}
c_{2}=\pm\varepsilon^{-1/4}/\left(\sqrt{2}\pi^{1/4}\right)\ .
\end{equation}

The all-time asymptotic approximation for the downsweep trajectory
is therefore
\begin{equation}
y_{\downarrow}\left(t\right)=\begin{cases}
{\displaystyle \frac{\pm\varepsilon^{1/4}e^{-t^{2}/2}}{\pi^{1/4}\sqrt{\textrm{erfc\ensuremath{\left(-t\right)}}}}} & t\le0\,,\ \text{or}\ t>0\ \text{and}\;e^{\frac{1}{2}t^{2}}\ll\varepsilon^{-1/4}\\
{\displaystyle \sqrt{\varepsilon}e^{-t^{2}/2}\biggl(\pm\frac{\varepsilon^{-1/4}}{\sqrt{2}\pi^{1/4}}-w\sqrt{\frac{\pi}{2}}\textrm{erfi}\Bigl(\frac{t}{\sqrt{2}}\Bigr)}\biggr) & t>0\ \text{and}\ 1\ll e^{\frac{1}{2}t^{2}}
\end{cases}\ ,\label{eq:downsweep}
\end{equation}
or
\begin{equation}
y_{\downarrow}\left(t\right)=\varepsilon^{1/4}\begin{cases}
{\displaystyle \frac{\pm e^{-t^{2}/2}}{\pi^{1/4}\sqrt{\textrm{erfc\ensuremath{\left(-t\right)}}}}} & t\le0\,,\ \text{or}\ t>0\ \text{and}\;e^{\frac{1}{2}t^{2}}\ll1/s\\
{\displaystyle e^{-t^{2}/2}\left(\pm\frac{1}{\sqrt{2}\pi^{1/4}}-s\sqrt{\frac{\pi}{2}}\textrm{erfi}\left(\frac{t}{\sqrt{2}}\right)\right)} & t>0\ \text{and}\ 1\ll e^{\frac{1}{2}t^{2}}
\end{cases}\ ,\label{eq:downsweep-s}
\end{equation}
in terms of the shape parameter. \\

\begin{figure}[t]
\includegraphics[scale=0.65]{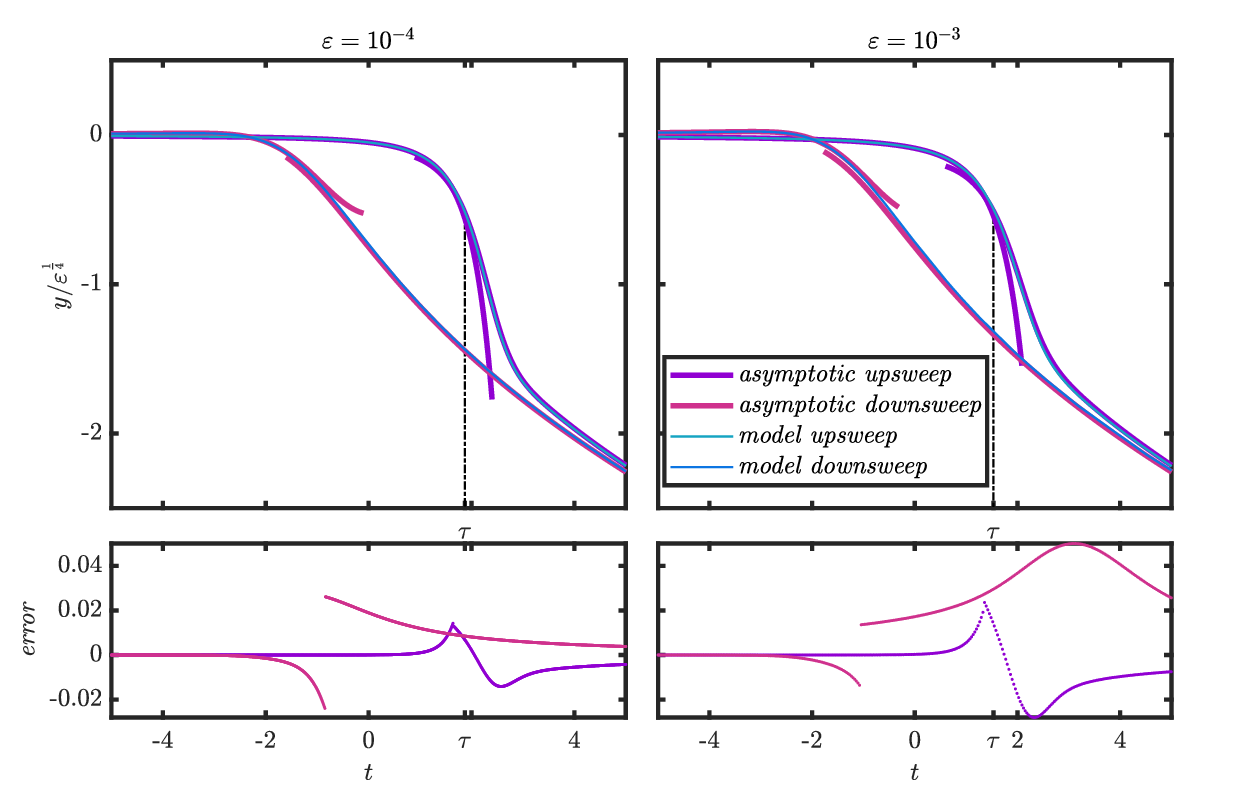}\caption{\label{fig:Asymptotic approximation and numerical solution}Top: The
asymptotic approximations of the adiabatic trajectories given in equations
(\ref{eq:upsweep-s}) and (\ref{eq:downsweep-s}) in purple (upsweep)
and violet (downsweep) overlayed with the numerically calculated continuum
model trajectories of equation (\ref{eq:nf-css}) in turquoise and
blue for upsweep and downsweep respectively. The asymptotic approximations
are valid in the intervals given in equations (\ref{eq:upsweep-s})
and (\ref{eq:downsweep-s}). Bottom: The difference between the asymptotic
approximation and the numerically calculated trajectories. In the
overlap interval where both the early- and late-time asymptotes are
valid, the smaller error in absolute value is shown. As in figure
(\ref{fig:Numerical model and map}) $\varepsilon=10^{-4}\,,\,w=0.4$
($s=0.04$) in the left panels, and $\varepsilon=10^{-3}\,,\,w=0.4$
($s=0.07$) in the right panels. As expected, the error in the asymptotic
approximations decreases with decreasing $\varepsilon$.}
\end{figure}

Figure \ref{fig:Asymptotic approximation and numerical solution}
shows a comparison between the asymptotic approximations given in
equations (\ref{eq:upsweep-s}) and (\ref{eq:downsweep-s}) and continuum
model numerical solutions of the (\ref{eq:nf-css}), confirming the
validity of the early and late asymptotic approximations in the expected
intervals, and the overlap of the validity intervals.

\subsection{The hysteresis loop\label{subsec:The-hysteresis-loop}}

The breakdown of adiabaticity near the bifurcation point breaks time
reversal symmetry, yielding essentially different dynamics in the
upsweep and downsweep directions. The most important expression of
this difference is the delay $\tau$ between the time that the control
parameter $r$ sweeps through the bifurcation point in the upward
direction, and the time where the map trajectory approaches the new
attractor at $\pm\varepsilon^{1/4}\sqrt{t}$.

A convenient definition for the upsweep breakout time $\tau$ is the
time where the shifted trajectory $y$ reaches the midway point $\pm\varepsilon^{1/4}\sqrt{t}/2$
between the instantaneous stable and unstable fixed points. For this
$t$ we can use the late-time asymptotic equation (\ref{eq:upsweep-s}),
approximating $\text{erfi}\left(t\right)\sim e^{t^{2}}/\left(\sqrt{\pi}t\right)$,
to obtain
\begin{equation}
\frac{\sqrt{\tau}}{2}=\frac{\sqrt{\tau}}{\sqrt{\left(2\pi s^{2}\right)^{-1}\tau e^{-\tau^{2}}+1}}\Rightarrow2\pi s^{2}\tau e^{\tau^{2}}=\frac{1}{3}\Rightarrow\tau\sim\sqrt{-\log\left(6\pi s^{2}\right)}
\end{equation}
in the adiabatic limit where $s\to0$; note that $\tau$ diverges
logarithmically in this limit.

The breakout time $\tau$ is a measure of the temporal width of the
upsweep-downsweep hysteresis loop, whose height is therefore $\sim\varepsilon^{1/4}\sqrt{\tau}$,
yielding the estimated area $\varepsilon^{1/4}\tau^{3/2}\sim\varepsilon^{1/4}\left(-\log s\right)^{3/4}$
for the area of the loop in the $y,t$ space, or $\varepsilon^{3/4}\left(-\log s\right)^{3/4}$
in the $x,r$ space.

To obtain a more precise approximation, we define the normalized loop
area
\begin{equation}
H=\frac{1}{\varepsilon^{1/4}}\left|\int_{-\infty}^{\infty}\left(y_{\downarrow}\left(-t\right)-y_{\uparrow}\left(t\right)\right)dt\right|
\end{equation}
that is a function of $s$ only. It follows from the preceding arguments
that $H$ diverges in the limit $s\to0$; our goal is to identify
the form of this divergence, and in this view we can subtract from
$H$ contributions that remain finite in this limit. Thus, we can
restrict the domain of integration to the interval $0\le t<\infty$,
and use the early--time approximation $\pm\varepsilon^{1/4}\sqrt{-t}$
for $y_{\downarrow}\left(t\right)$ and the late-time approximation
$\pm\varepsilon^{1/4}\sqrt{t/\left(1+\left(2\pi s^{2}\right)^{-1}te^{-t^{2}}\right)}$
for $y_{\uparrow}\left(t\right)$, so that as $s\to0$,
\begin{equation}
H\sim\int_{0}^{\infty}\sqrt{t}\left(1-\frac{1}{\sqrt{1+\left(2\pi s^{2}\right)^{-1}te^{-t^{2}}}}\right)\,dt\ .
\end{equation}
As observed above, for $t\lesssim\tau$, the second term in the integrand
is much smaller than the first, but for $t\gg\tau$ the two terms
in the integrand cancel, so that the integration is effectively cut-off
at $\tau$ from above. An analysis of the asymptotic approximation
of the area, outlined in appendix (\ref{sec:ha}), shows that in the
limit $s\to0$
\begin{equation}
H\mathop{{\sim}}\frac{2}{3}\left(\log\left(\frac{\sqrt{-\log\left(2\pi s^{2}\right)}}{2\pi s^{2}}\right)\right)^{3/4}+H_{0}\ ,\label{eq:has}
\end{equation}
where $H_{0}$ is an indeterminate constant; least squares fitting
yields $H_{0}\approx0.8853$ \\

\begin{figure}
\includegraphics[scale=0.65]{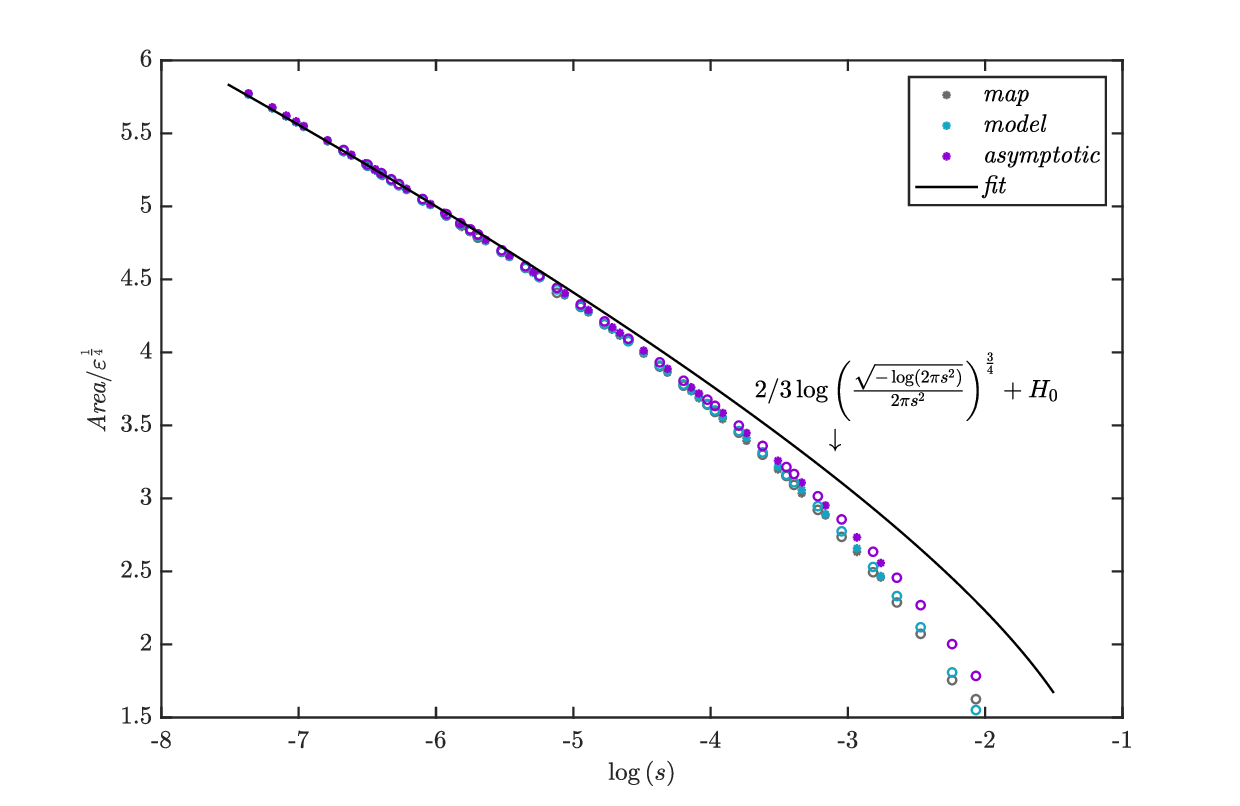}

\caption{\label{fig:Area vs epsilon}The area enclosed within the hysteresis
loop in the $y,t$ space divided by $\varepsilon^{1/4}$ versus the
natural logarithm of the shape parameter $s$. The area is calculated
in four different ways: using the adiabatic trajectories of the normal
form map (grey), the numerically calculated trajectories of the continuum
model equation (turquoise), the asymptotic approximations of the continuum
trajectories (purple), and the explicit asymptotic approximation of
the area (\ref{eq:has}) (black curve). The map trajectories were
calculated for $w=0.4$ (circles) and $w=02$ (asterisk).}
\end{figure}

Figure \ref{fig:Area vs epsilon} compares $H\left(s\right)$, calculated
with the numerical solutions of (\ref{eq:nf-css}) and with the asymptotic
trajectories (\ref{eq:upsweep-s}) and (\ref{eq:downsweep-s}), with
the asymptote (\ref{eq:has}) on one hand, and with the area of the
hysteresis loop of the map trajectories divided by $\varepsilon^{1/4}$
on the other hand, showing agreement between all four measures of
the area, that improves as $s$ decreases.

\subsection{Universality of the adiabatic trajectories: the logistic map\label{subsec:Logistic-map}}

As argued above, the adiabatic sweep trajectories that we derived
and the resultant hysteresis loop are universal; they capture the
shape of any sufficiently slow sweep trajectories of a map through
a period-doubling or pitchfork bifurcation, sufficiently close to
the bifurcation point. We next demonstrate this claim for the case
of the logistic map 
\begin{equation}
x_{n+1}=L_{n}\left(x_{n}\right)\ ,\qquad L_{n}\left(x\right)=\left(3\pm\frac{1}{2}\varepsilon n\right)x\left(1-x\right)\ 
\end{equation}
with $\varepsilon>0$ small. Note that we parametrize $L_{n}\left(x\right)$
such that it passes through the fundamental period-doubling bifurcation
of the logistic map at $n=0$. As discussed above, the universal trajectories
in equations (\ref{eq:upsweep}) and (\ref{eq:downsweep}) are adiabatic
asymptotic approximants to trajectories of $M_{n}=L_{n}^{2}$, which
undergoes a supercritical pitchfork bifurcation at $n=0$; meanwhile,
trajectories of $M_{n}$ are adiabatic approximants of trajectories
of $L_{n}\circ L_{n-1}$.

To bring the second iterate of the logistic map to normal form, we
apply a linear transformation, shifting the bifurcation point to zero
and stretching the map vertically so that the positive-$n$ outer
instantaneous fixed points of the map are at the standard $\pm\sqrt{\varepsilon n}+O\left(\varepsilon\right)$.
This produces the stretched and shifted map $N_{n}=\sqrt{18}\left(M_{n}\left(\frac{x}{\sqrt{18}}+\frac{2}{3}\right)-\frac{2}{3}\right)$;
the instantaneous inner fixed point of $N_{n}$ is at $\varepsilon n/\sqrt{18}+O\left(\varepsilon^{2}\right)$,
and thus $w=1/\sqrt{18}$ in this example.\\

\begin{figure}
\includegraphics[scale=0.65]{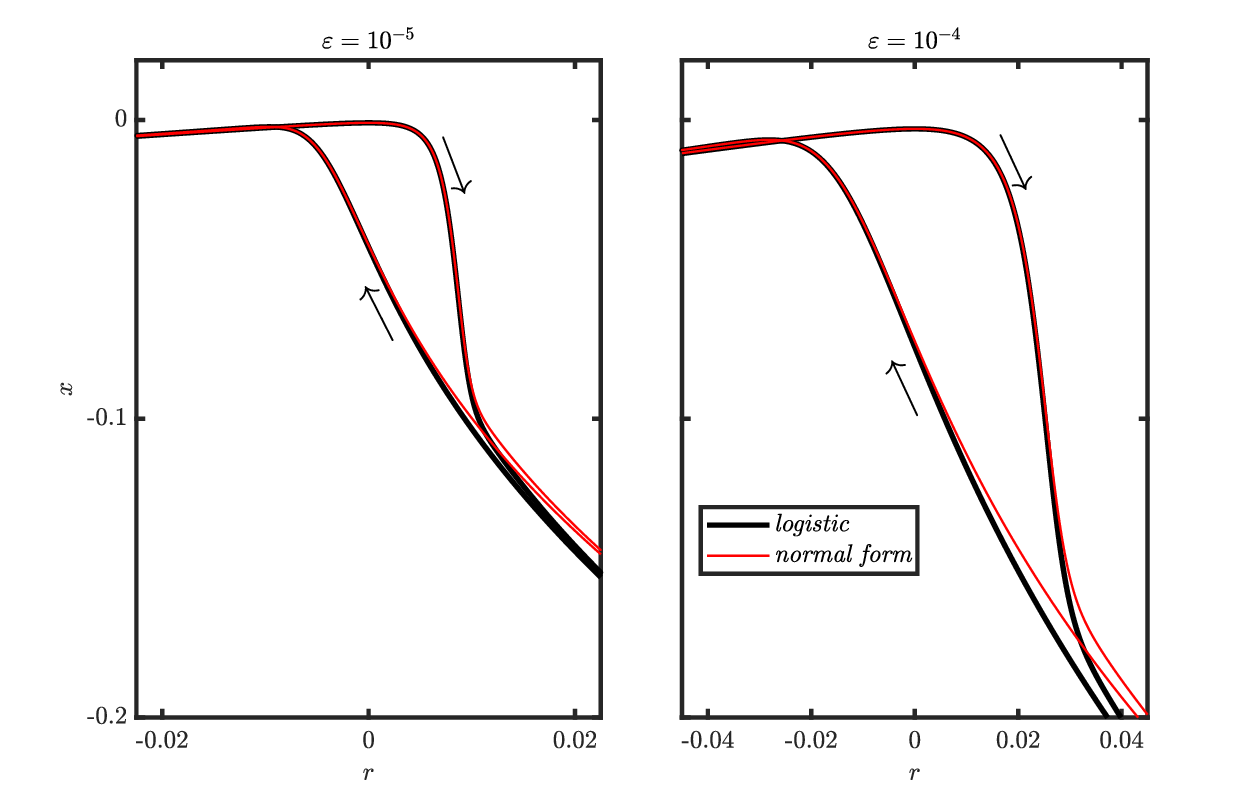}

\caption{\label{fig:Logistic_in_normal_form}Adiabatic sweep trajectories of
the shifted and stretched second iterate of the logistic map $N_{n}$
(black) and of the normal map $F_{n}$ (red), for two values of the
adiabatic parameter (sweep rate) $\varepsilon$. The agreement is
good and improves for decreasing $\varepsilon$.}
\end{figure}

Figure \ref{fig:Logistic_in_normal_form} shows a comparison between
the adiabatic sweep trajectories of $N_{n}$ and those of the the
normal form map $F_{n}$ of equation (\ref{eq:pitchfork}). The figure
confirms that for small enough $r$ and $\varepsilon$, the the adaiabtic
logistic map indeed behaves like the universal map, with $\varepsilon$
of the first being half the $\varepsilon$ of the second.

\selectlanguage{british}%

\section{Discussion and Conclusions \label{sec:Discussion-and-Conclusions}}

\selectlanguage{english}%
Bifurcations in autonomous dynamical systems are important because
they mark a qualitative change in the long-term behavior of the system.
The study of bifurcations is facilitated by the transformation to
a normal form, which allows one to calculate the trajectories of the
system near the bifurcation. Since the normal form has a universal
structure, the dynamics near the bifurcation are universal.

Here we analyzed the dynamics of an adiabatically time-dependent map
that is swept past a period-doubling or pitchfork bifurcation. The
first main result of this analysis is the development of the adiabatic
normal form for this class of bifurcations from which it follows that
the universal adiabatic sweep dynamics depends on a structural asymmetry
parameter, which is irrelevant for the autonomous bifurcation analysis.
The shape of the adiabatic sweep trajectories is determined, up to
scaling, by the shape parameter, a combination of the asymmetry parameter
and the adiabatic small parameter (which is the sweep rate). Thus,
even though all autonomous period-doubling and pitchfork bifurcations
have the same structure after appropriate scaling, there is a one-parameter
family of structures of adiabatic sweeping trajectories of these bifurcations.

Our second main result is the complete asymptotics of the up- and
downsweep adiabatic trajectories, explicitly expressed in terms of
error functions. In both directions, the trajectory initially follows
closely one of the instantaneous stable fixed points, but opens an
increasing gap from it upon approaching the bifurcation, before finally
breaking back toward stable fixed points after leaving the bifurcation
region. The breakdown of adiabaticity at the bifurcation is the source
of asymmetry between the up- and downsweep trajectories that gives
rise to hysteresis.

On the basis of these explicit asymptotic expressions, we next calculated
the area of the hysteresis loop enclosed between the upsweep and downsweep
trajectories. We show that even though the adiabatic-trajectory shape
parameter tends to zero in the adiabatic limit, the hysteresis loop
itself does not converge to a well-defined shape in this limit, and
as a consequence, the area of the loop has a complicated logarithmic
adiabatic asymptote. Finally, as an application of our results, we
showed that the universal sweep trajectories correctly capture the
dynamics of the logistic map with an adiabatically varying parameter
in the region of the fundamental period-doubling bifurcation.

The results of this work are geared to be useful for the analysis
of adiabatic sweep experiments and simulations. While the breakdown
of adiabaticity implies that the autonomous dynamics close to a bifurcation,
and in particular the location of the bifurcation point itself, can
never be faithfully reproduced by a finite speed sweeping, full information
about the bifurcation structure can be extracted from the sweep trajectories.
Namely, a fit of the experimental trajectories with the universal
ones would yield both the location of the bifurcation point and the
values of the map parameters at the bifurcation, from which it is
possible to reconstruct the bifurcation diagram of the autonomous
map.

\appendix

\section{Asymptotic approximation of the area of the hysteresis loop\label{sec:ha}}

In this appendix we derive an asymptotic approximation for the integral
\begin{equation}
H_{a}=\int_{0}^{\infty}\sqrt{t}\left(1-\frac{1}{\sqrt{1+ate^{-t^{2}}}}\right)\,dt\ =\frac{2}{3}\int_{0}^{\infty}\left(1-\frac{1}{\sqrt{1+at^{2/3}e^{-t^{4/3}}}}\right)\,dt\ 
\end{equation}
 in the limit $a=\left(2\pi s^{2}\right)^{-1}\to\infty$. Changing
variable to $x=e^{-t^{4/3}}$ we get
\begin{equation}
H_{a}=\frac{1}{2}\int_{0}^{1}\left(1-\frac{1}{\sqrt{1+ax\sqrt{-\log x}}}\right)\,\frac{dx}{x\left(-\log x\right)^{1/4}}=\int_{0}^{1}\frac{a\left(-\log x\right)^{1/4}dx}{2\sqrt{1+ax\sqrt{-\log x}}\left(\sqrt{1+ax\sqrt{-\log x}}+1\right)}\ .\label{eq:ha}
\end{equation}
To estimate this integal, note that for a given $a\gg1$, and $x$
which is not too small, $ax\sqrt{-\log x}\gg1$, and for such $x$
\begin{equation}
\frac{a\left(-\log x\right)^{1/4}}{2\sqrt{1+ax\sqrt{-\log x}}\left(\sqrt{1+ax\sqrt{-\log x}}+1\right)}\sim\frac{1}{2x\left(-\log x\right)^{1/4}}\ ,\label{eq:largex}
\end{equation}
which is logarithmically divergent for small $x$.

On the other hand for sufficiently small $x$, $ax\sqrt{-\log x}\ll1$,
and for such $x$, the integrand in equation (\ref{eq:ha}) is close
to $a(-\log x)^{1/4}/2$, which is integrable at 0. The crossover
between the two limits occurs for $x\sim1/\left(a\sqrt{\log a}\right)$,
so it is advantageous to express
\begin{equation}
H_{a}=\left(\int_{0}^{k/\left(a\sqrt{\log a}\right)}+\int_{k/\left(a\sqrt{\log a}\right)}^{1}\right)\frac{a\left(-\log x\right)^{1/4}dx}{2\sqrt{1+ax\sqrt{-\log x}}\left(\sqrt{1+ax\sqrt{-\log x}}+1\right)}\ ,\label{eq:ha-1}
\end{equation}
with $k$ an order-one number. The small-$x$ integral is $\sim\frac{k}{4}\left(\log a\right)^{-1/2}\left(\log\left(a\sqrt{\log a}/k\right)\right)^{1/4}$,
so can be neglected in the limit $a\to\infty$, and we can let $k$
be large enough that the estimate (\ref{eq:largex}) is valid for
all $x$ in the larger-$x$ integral, obtaining finally
\begin{equation}
H_{a}\sim\int_{k/\left(a\sqrt{\log a}\right)}^{1}\frac{1}{2x\left(-\log x\right)^{1/4}}\sim\frac{2}{3}\left(\log\left(a\sqrt{\log a}\right)\right)^{3/4}
\end{equation}
\\

\bibliographystyle{apsrev4-2}
\bibliography{Hysteresis}

\end{document}